\documentclass[twocolumn,prstab,showpacs]{revtex4}

\usepackage{amsmath}
\usepackage{subfigure}

\usepackage{tikz}
\usepackage{pgfplots}

\begin{document}

	\title{Electron beam profile imaging in the presence of coherent optical radiation effects}

	\author{Christopher Behrens$^\mathrm{1}$, Christopher Gerth$^\mathrm{1}$, Gero Kube$^\mathrm{1}$, Bernhard Schmidt$^\mathrm{1}$, Stephan Wesch$^\mathrm{1}$, and Minjie Yan$^\mathrm{1,2}$}
	\affiliation{$^\mathrm{1}$ Deutsches Elektronen-Synchrotron DESY, Notkestr.\,85, 22607 Hamburg, Germany\\
	\mbox{$^\mathrm{2}$ Universit\"at Hamburg, Institut f\"ur Experimentalphysik, Luruper Chaussee 149, 22761 Hamburg, Germany}}
	\date{\today}	

	\begin{abstract}
	High-brightness electron beams with low energy spread at existing and future x-ray free-electron lasers are affected by various collective beam self-interactions and microbunching instabilities. The corresponding coherent optical radiation effects, e.g., coherent optical transition radiation, impede electron beam profile imaging and become a serious issue for all kinds of electron beam diagnostics using imaging screens. Furthermore, coherent optical radiation effects can also be related to intrinsically ultrashort electron bunches or the existence of ultrashort spikes inside the electron bunches. In this paper, we discuss methods to suppress coherent optical radiation effects both by electron beam profile imaging in dispersive beamlines and by using scintillation imaging screens in combination with separation techniques. The suppression of coherent optical emission in dispersive beamlines is shown by analytical calculations, numerical simulations, and measurements. Transverse and longitudinal electron beam profile measurements in the presence of coherent optical radiation effects in non-dispersive beamlines are demonstrated by applying a temporal separation technique.

	\end{abstract}

	\pacs{29.27.-a, 41.60.Cr, 41.60.Dk}
    	
	\maketitle

	\section{Introduction}
	X-ray free-electron lasers (FELs) offer a brilliant tool for science at atomic length and ultrafast time scales~\cite{LCLSnature1}, and they have been realized with the operation of the Free-Electron Laser in Hamburg (FLASH)~\cite{FLASHAnature}, the Linac Coherent Light Source (LCLS)~\cite{LCLSnature2}, and the SPring-8 Angstrom Compact Free Electron Laser (SACLA)~\cite{SACLAnature}. The x-ray FEL driving electron bunches are subject to several collective effects, e.g., microbunching instabilities or coherent synchrotron radiation (CSR), which degrade the required high transverse and longitudinal beam brightness~\cite{CSR,CSR-ub,lsc-ub,Venturini}. These instabilities may not only result in significant deteriorations of the FEL performance~\cite{LCLSheater} but also in coherent radiation effects~\cite{Glin,loos,linLCLS,sears,lump1,lump2,Wesch1} such as coherent optical transition radiation (COTR) or CSR in the optical wavelength range~\cite{bane} (abbreviated as COSR). Beam profile imaging dominated by coherent optical radiation leads to an incorrect representation of the transverse charge distribution~\cite{loos} and renders electron beam diagnostics with standard imaging screens, e.g., OTR screens, and all the related diagnostics such as emittance or bunch length diagnostics impossible. However, beam diagnostics with imaging screens are essential for single-shot measurements or in cases where two transverse dimensions are required, e.g., in slice-emittance or longitudinal phase space measurements~\cite{Roehrs,Filippetto,pulse}.
	
	Microbunching instabilities associated with longitudinal electron bunch compression can be mitigated by introducing additional uncorrelated energy spread~\cite{emma,LH,TH} as successfully demonstrated by the operation of the laser heater system at the LCLS~\cite{LCLSheater}. However, the microbunching gain suppression is not necessarily perfect, and the corresponding remaining small but existing level of COTR still hampers electron beam profile diagnostics using standard imaging screens (e.g., Ref.~\cite{LCLSheater}). The origin of coherent optical radiation effects is not only restricted to microbunching instabilities but can also be related to ultrashort spikes inside electron bunches or generated by intrinsically ultrashort electron bunches like at laser-plasma accelerators (e.g., Ref.~\cite{short}) or at x-ray FELs with ultra-low charge operation~\cite{Ding,Aline2,Igor}.

	Transition radiation is emitted when a charged particle beam crosses the boundary between two media with different dielectric properties~\cite{tr1,tr2,tr3,hon,dao}, hence transition radiation is emitted using any kind of imaging screen and thus precludes the stand-alone use of scintillation screens in the presence of coherent optical radiation effects (e.g., COTR). However, by using (scintillation) imaging screens in dedicated measurement configurations, COTR can be mitigated (see, e.g., Ref.~\cite{lump2}). 

	In this paper, we discuss methods to suppress coherent optical radiation effects both by electron beam profile imaging in dispersive beamlines and by utilizing scintillation imaging screens in combination with several separation techniques. The experimental setup and observations of coherent optical radiation effects at FLASH are described in Sec.~\ref{sec:setup}. In Sec.~\ref{sec:ES} we discuss the suppression of coherent optical emission in dispersive beamlines and present experimental results for COTR generated by a local ultrashort charge concentration. Section~\ref{sec:sep} covers the suppression of coherent optical radiation effects by using scintillation screens in combination with separation techniques. The experimental results obtained with the temporal separation technique are presented in Sec.~\ref{sec:res}, and a summary and conclusions are given in Sec.~\ref{sec:Summary}.

	\section{Experimental setup and observation of coherent effects}\label{sec:setup}

	\begin{figure*}[htb]
		\centering
		\includegraphics[width=0.9\linewidth]{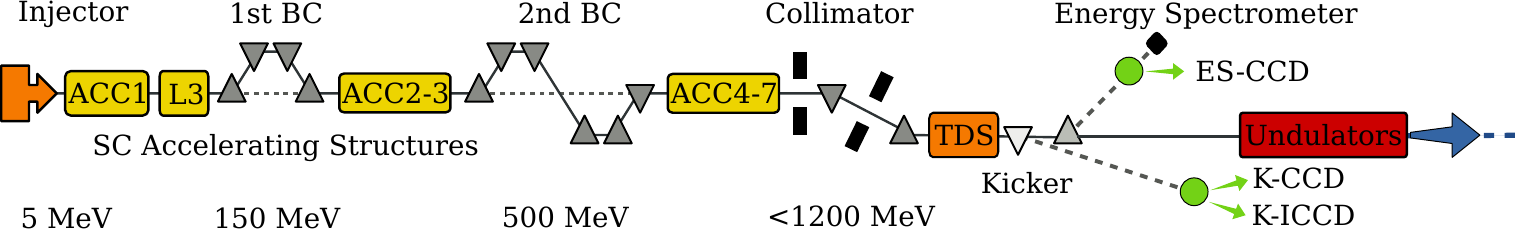} 
		\caption{Schematic layout of the Free-Electron Laser in Hamburg (FLASH) with its superconducting (SC) accelerating structures (ACC), the two magnetic bunch compressor (BC) chicanes, and the third-harmonic rf linearizer system (L3). The positions of the experimental setups and diagnostics used for the measurements presented in this paper are indicated by green dots.}
		\label{fig:FLASH_1}
	\end{figure*}

	The measurements presented in this paper have been carried out at FLASH, which is a self-amplified spontaneous emission (SASE) FEL~\cite{kon} for extreme-ultraviolet (EUV) and soft x-ray radiation, driven by a superconducting radio-frequency (rf) linear accelerator~\cite{FLASHAnature}. The schematic layout of FLASH is depicted in Fig.~\ref{fig:FLASH_1}, showing the injector, which is based on a laser-driven normal conducting rf gun, the superconducting accelerating structures, two magnetic bunch compressor chicanes, and the undulator magnet system. The positions of the experimental setups used for the measurements presented in this paper are indicated by green dots and arrows. 

	The third-harmonic rf system (denoted by L3 in Fig.~\ref{fig:FLASH_1}) is dedicated to the linearization of the longitudinal phase space upstream of the first bunch compressor~\cite{pulse,lin2}. In order to properly set up FEL operation with applied third-harmonic rf linearizer, a LOLA-type~\cite{LOLA} transverse deflecting rf structure (TDS) has been integrated in a dedicated setup for diagnosis of the longitudinal phase space~\cite{LOLA1,LOLA2} close to the FEL undulators. As depicted in Fig.~\ref{fig:FLASH_1}, the TDS can either be operated in combination with imaging screens in the dispersive magnetic energy spectrometer or by using off-axis imaging screens operated with a fast kicker magnet in the non-dispersive main beamline during FEL operation. Technical details and performance measurements on the setup for longitudinal beam diagnostics can be found in Refs.~\cite{pulse,LOLA1,LOLA2}.

	\subsection{Time-domain longitudinal beam diagnostics}\label{subsec:lola}
	 Transverse deflecting rf structures are widely used for electron bunch length and longitudinal profile measurements at present FELs and provide high-resolution single-shot diagnostics~\cite{Roehrs,Filippetto,pulse,Kick}. Detailed descriptions of time-domain electron bunch diagnostics using a TDS can be found in Refs.~\cite{Roehrs,Kick}. Here we describe only the basic principles of longitudinal electron beam diagnostics that are required throughout this paper.

		The vertical betatron motion of an electron passing a vertical deflecting TDS around the zero-crossing rf phase, neglecting intrinsic longitudinal-to-vertical correlations~\cite{pulse} which are not relevant for the experiments presented throughout this paper, can be given by~\cite{Roehrs,pulse}\begin{equation}
			y(s) =  y_0(s) + S_y(s,s_0)c^{-1}z(s_0)
		\label{eq:motion}
		\end{equation} with the vertical shear (streak) function
		\begin{equation}
			S_y(s,s_0) = R_{34}K_y= \sqrt{\beta_y(s)\beta_y(s_0)}\mathrm{sin}(\Delta\phi_y) \frac{e \omega V_y}{pc}\,,
		\label{eq:beta}
		\end{equation} where $R_{34}=\sqrt{\beta_y(s)\beta_y(s_0)}\mathrm{sin}(\Delta\phi_y)$ is the angular-to-spatial element of the vertical beam transfer matrix from the TDS at $s_0$ to any position $s$, $\beta_y$ is the vertical beta function, $\Delta\phi_y$ is the vertical phase advance between $s_0$ and $s$, and $y_0$ describes an intrinsic offset. The expression $K_y = e \omega V_y/(pc)$ is the vertical kick strength with the peak deflection voltage $V_y$ in the TDS, $c$ is the speed of light in vacuum, $e$ is the elementary charge, $p$ is the electron momentum, $z(s_0)$ is the longitudinal position of the electron relative to the zero-crossing rf phase, and $\omega/(2\pi)$ is the operating rf frequency. The expression in Eq.~(\ref{eq:motion}) shows a linear mapping from the longitudinal to the vertical coordinate and allows longitudinal electron beam profile measurements by means of transverse beam diagnostics using imaging screens. The shear function $S_y$ determines the slope of this mapping and can be calibrated by measuring the vertical centroid offset of the bunch as a function of the TDS rf phase. The electron bunch current is given by the normalized longitudinal bunch profile multiplied by the electron bunch charge. The bunch length (duration) is given by the root mean square (r.m.s.) value $\sigma_{t,\mathrm{e}} = S^{-1} (\sigma_y^2 - \sigma_{y,0}^2)^{1/2}$, where $\sigma_y$ is the vertical r.m.s.~beam size during TDS operation, and $\sigma_{y,0}$ is the intrinsic vertical r.m.s.~beam size when the TDS is switched off. Both $\sigma_y$ and $\sigma_{y,0}$ can be determined by measurements, and the latter limits the achievable r.m.s.~time resolution to $\mathcal{R}_{t,\mathrm{e}}=\sigma_{y,0}/S_y$~\cite{Roehrs,pulse}.
	
	\subsection{Imaging screen stations and camera systems}\label{sec:screens}

	The screen stations in both the magnetic energy spectrometer and non-dispersive main beamline (see Fig.~\ref{fig:FLASH_1}) are each equipped with different imaging screens and a charge-coupled device (CCD) camera~\cite{pro} (1360$\times$1024 pixels with 12\,bit dynamic range and $6.45\times6.45\,\mathrm{\mu m}^2$ pixel size) with motorized optics (motorized macro lens with teleconverter mounted on a linear translation stage). The translation stage allows variable demagnification $M^{-1}$ in the range between $\sim$\,1.5 - 3 with spatial resolutions of better than $16\,\mathrm{\mu m}$. The imaging screen station in the energy spectrometer (ES-CCD in Fig.~\ref{fig:FLASH_1}) is equipped with an OTR screen (aluminum coated silicon) and two scintillation screens made of cerium-doped yttrium aluminum garnet (YAG:Ce) and bismuth germanate (BGO), respectively. In the non-dispersive beamline, the screen station is operated with a fast kicker magnet (K-CCD in Fig.~\ref{fig:FLASH_1}), which is able to deflect one bunch out of the bunch train at the bunch train repetition rate of FLASH~\cite{flash} of 10\,Hz, and provides an OTR screen and a cerium-doped lutetium aluminum garnet (LuAG:Ce) scintillation screen. All screens are mounted at a 45$^\circ$ angle (the cameras at a 90$^\circ$ angle) with respect to the incoming electron beam. The scintillation screens have a thickness of $100\,\mathrm{\mu m}$. The experimental setup in the non-dispersive beamline is additionally equipped with a fast gated intensified CCD camera~\cite{pco} (K-ICCD in Fig.~\ref{fig:FLASH_1}, 1280$\times$1024 pixels with 12\,bit and $6.7\times6.7\,\mathrm{\mu m}^2$ pixel size), which has been used for the temporal separation technique (see Sec.~\ref{sec:res}). Further technical details on the screen stations and camera systems can be found in Refs.~\cite{LOLA2,yan1}.

	\subsection{Observation of coherent optical transition radiation and microbunching in the time-domain}\label{sec:subsecCOTR}
	Microbunching instabilities at x-ray FELs can lead to significant generation and amplification of density modulations in the optical wavelength range~\cite{CSR,CSR-ub,lsc-ub} which may result in coherent optical radiation effects such as COTR. This has been observed by spectral measurements and characteristic ring-shaped light patterns at the LCLS~\cite{loos,linLCLS} and FLASH~\cite{Wesch1}, and renders accurate electron beam profile diagnostics using standard imaging screens impossible. First observations of COTR~\cite{Wesch1} and microbunching in the frequency-domain (coherent transition radiation around 10 $\mathrm{\mu m}$~\cite{Wesch2}) at FLASH were made directly upstream of the collimator (see Fig.~\ref{fig:FLASH_1}).
	\begin{figure}[b]
	\centering
    \subfigure[~OTR screen.]{\includegraphics[width=0.48\linewidth]{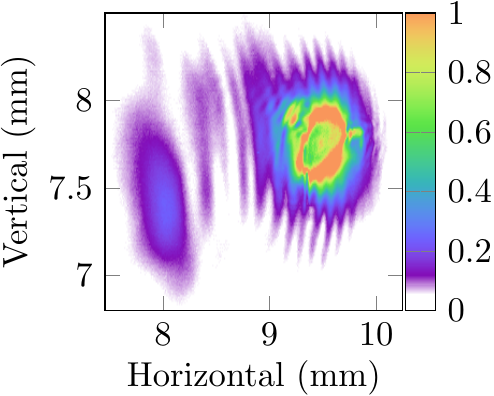} \label{fig:Indi_2_a}}
	\subfigure[~LuAG screen.]{\includegraphics[width=0.48\linewidth]{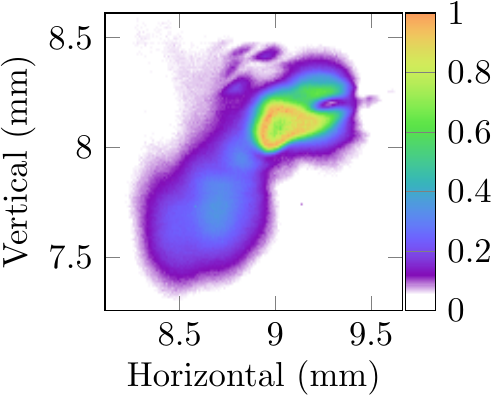} \label{fig:Indi_2_b}}
	\caption{Single-shot images of light patterns at the imaging screens (K-CCD) generated by compressed electron bunches: (a) OTR screen and (b) LuAG screen. For both images a long-pass filter, blocking wavelengths below 780\,nm, was used.}
	\label{fig:Indi_2}
	\end{figure} 
	\begin{figure}[t]
	\centering
    \subfigure{\includegraphics[width=0.48\linewidth]{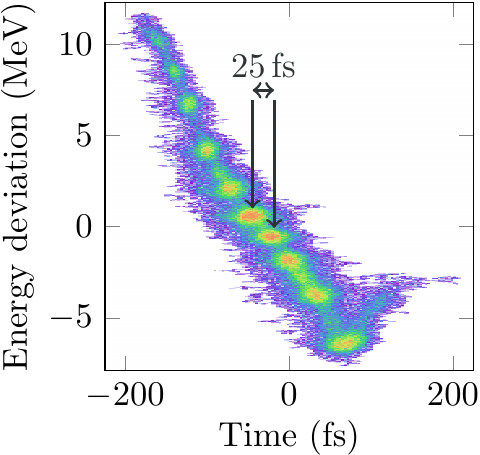} \label{fig:Indi_3_a}}
	\subfigure{\includegraphics[width=0.48\linewidth]{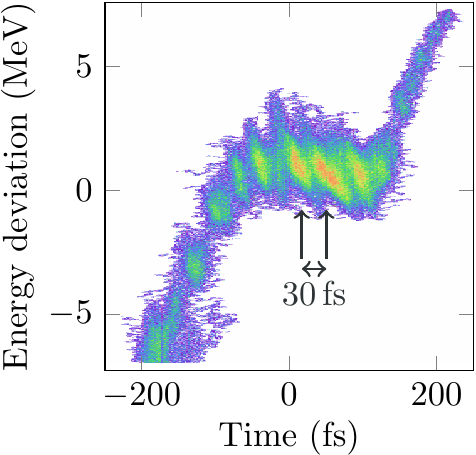} \label{fig:Indi_3_b}}
	\caption{Longitudinal phase space measurements upstream of the undulators at ES-CCD for two different compression settings and mean energies: (a) 796\,MeV and (b) 661\,MeV. The density modulations indicate microbunching in the time-domain with periods of $\sim25$\,fs and 30\,fs, respectively.}
	\label{fig:Indi_3}
	\end{figure}Electron beam profile imaging performed downstream of the collimator section~\cite{LOLA2}, an achromatic bending system, resulted in considerably more prominent observation of coherent optical radiation effects and microbunching.
	
	The measurements presented in Fig.~\ref{fig:Indi_2} show single-shot light patterns, generated by moderately compressed electron bunches, at the imaging screens in the non-dispersive main beamline at K-CCD directly upstream of the undulators. Ring-shaped structures in the profiles, characteristic for COTR~\cite{loos}, are clearly visible in the images of Figs.~\ref{fig:Indi_2_a} and~\ref{fig:Indi_2_b}, which have been recorded by using an OTR and LuAG imaging screen, respectively. For both images a long-pass filter, blocking wavelengths below 780\,nm, was used. The luminescence emission of the LuAG scintillation screen occurs below 700\,nm~\cite{bla} and is thus well blocked by the 780\,-nm long-pass filter used during the measurements. Hence, the light pattern in Fig.~\ref{fig:Indi_2_b} is due to COTR without contribution from scintillation light. Complementary to the observation of COTR, the images in Fig.~\ref{fig:Indi_3} show single-shot longitudinal phase space measurements in the magnetic energy spectrometer (ES-CCD). The measurements were done for accelerator settings typical for FEL operation with applied third-harmonic rf linearizer system upstream of the bunch compressor chicanes, and they clearly indicate microbunching in the time-domain with modulation periods of about 25\,fs and 30\,fs, respectively. We note that a maximum modulation wavelength of $10\,\mathrm{\mu m}$ ($33\,\mathrm{fs}$) was predicted theoretically in Ref.~\cite{lsc-ub} and measured by spectroscopy of coherent transition radiation in Ref.~\cite{Wesch2}.

	\section{Suppression of coherent optical emission in dispersive beamlines}\label{sec:ES}
	The energy-dependent beam trajectories in dispersive beamlines can be utilized as a magnetic energy spectrometer for charged particle beams. By combining such an energy spectrometer with the operation of a TDS and using imaging screens to get two-dimensional transverse beam profiles, longitudinal phase space measurements (see, e.g., Fig.~\ref{fig:Indi_3}) with single-shot capability can be accomplished. The corresponding horizontal betatron motion, which should be perpendicular to the vertical shearing plane of the TDS~\cite{Roehrs,pulse}, can be written as 
	\begin{equation}
			x(s) =  x_0(s) + D_x(s,s_0)\delta(s_0)
	\label{eq:disp}
	\end{equation} with the intrinsic offset $x_0$, the horizontal momentum dispersion $D_x(s,s_0)$ and the relative momentum deviation $\delta = \Delta p/p$. For relativistic electron beams with Lorentz factors of $\gamma \gg 1$, the electron beam energy is given by $E\approx pc$, and $\delta$ represents the relative energy deviation. 

	The dispersion $D_x$ can be determined by measuring the horizontal centroid offset of the bunch as a function of the relative energy deviation. The dispersion in the magnetic energy spectrometer at ES-CCD (see Fig.~\ref{fig:FLASH_1}), which is generated by two subsequent dipole magnets with 5$^\circ$ deflection each (equivalent to a single dipole magnet with 10$^\circ$ deflection), amounts to 750\,mm (nominal)~\cite{pulse}, whereas $D_x$ at K-(I)CCD due to the kicker magnet operation is negligible. In addition to the momentum dispersion introduced in the horizontal betatron motion, the longitudinal particle motion can be described by
	\begin{equation}
			z(s) =  z(s_0) + R_{51}x(s_0) + R_{52}x'(s_0) + R_{56}\delta(s_0)
	\label{eq:longdisp}
	\end{equation}
	with the initial bunch length coordinate $z$ and the initial horizontal offset $x$ and slope $x'=dx/ds$. The transfer matrix elements $R_{ij}$ describe the mapping from position $s_0$ to $s$, i.e., $R_{ij}\equiv R_{ij}(s_,s_0)$ throughout the rest of this paper. The expression in Eq.~(\ref{eq:longdisp}) does not affect the principle of longitudinal phase space diagnostics described by Eqs.~(\ref{eq:motion}) and~(\ref{eq:disp}), but results in the suppression of coherent optical emission as is shown in the following.

		\subsection{Analytical calculations and numerical particle tracking simulations}\label{sec:supp}
		 The spectral and angular intensity distribution, denoted as $\mathcal{I}(\vec{k})\equiv dI(\vec{k})/d\Omega$ with the three-dimensional wave vector $\vec{k} = (\vec{k}_r,k_z)$, of transition (synchrotron) radiation emitted by an electron bunch with $N\gg1$ electrons and charge $Q=Ne$ is given by (e.g., Refs.~\cite{Grimm,Wesch3})
		\begin{equation}
			\mathcal{I}(\vec{k}_r,k_z) =N \mathcal{I}_1(\vec{k}_r,k_z) + N^2 |F(\vec{k}_r,k_z)|^2 \mathcal{I}_1(\vec{k}_r,k_z)\,,
		\label{eq:spe}
		\end{equation}
		where $\mathcal{I}_1(\vec{k}_r,k_z)$ describes the intensity distribution of a single electron as a function of the transverse and longitudinal wavenumber $\vec{k}_r$ and $k_z$, respectively, and $F(\vec{k}_r,k_z)$ is the three-dimensional form factor of the electron bunch. The latter can be expressed by the Fourier transform of the normalized charge density $\rho(\vec r\,,z)$ as
		\begin{equation}
			F(\vec{k}_r,k_z)= \int{d\vec{r}\,dz\,\rho(\vec{r}\,,z)e^{-i \vec{k}_r\vec{r} }e^{-i k_zz } }\,,
		\label{eq:form}
		\end{equation}
		where $\rho(\vec r\,,z)\equiv \rho(x,y,z)$. Normalized charge distributions without longitudinal-transverse correlations can be factorized as $\rho(\vec r,z)\equiv \rho(\vec r\,)\rho(z)$, and by taking into account $\int{d\vec{r}\,\rho(\vec{r}\,)}=\int{dz\,\rho(z)}=1$, which is assumed in the following, we get $F(\vec{k}_r,k_z)=F_t(\vec{k}_r)\,F_l(k_z)$ with the transverse and longitudinal form factor $F_t$ and $F_l$, respectively. For small observation angles $\theta$ (small covered solid angles $\Omega$) with respect to the central axis ($z$-axis) of the emitted radiation we have $k_z = k \cos\theta \approx k$ with the wavenumber $k$, and the expression in Eq.~(\ref{eq:spe}) reads
		\begin{equation}
			\mathcal{I}(k,\Omega) \approx N \mathcal{I}_1(k,\Omega) + N^2 |F_l(k)|^2 |F_t(k,\Omega)|^2  \mathcal{I}_1(k,\Omega)
		\label{eq:spec}
		\end{equation}
		 The first term on the right-hand side is linear in $N$ and describes the contribution of incoherent radiation, whereas the second term scales with $N^2|F_l(k)|^2|F_t(k,\Omega)|^2$, which describes the coherent radiation part. In order to perform electron beam diagnostics with incoherent radiation, we demand that the total spectral radiation intensity in Eq.~(\ref{eq:spec}) is dominated by the incoherent term, i.e., $N \gg N^2|F_l(k)|^2|F_t(k,\Omega)|^2$. 

		In following, we derive an analytical expression describing a general strong suppression of the longitudinal form factor at optical wavelengths in a magnetic energy spectrometer. A transverse form factor of $|F_t|=1$, i.e., full transverse coherence, at the imaging screens is assumed, which is the worst case scenario. The actual transverse form factor in the experiment will be reduced due to the finite beam size and observation angle~\cite{Grimm}. However, the suppression of the longitudinal form factor $F_l$ presented below is much stronger in the general case. A cutoff wavelength $\lambda_{c}=2\pi/k_c$ can be defined via $|F_l(k_c)| = N^{-1/2}$, and beam diagnostics at wavelengths below $\lambda_{c}$ becomes dominated by incoherent radiation. The cutoff wavelength initially depends on the charge distribution [via Eq.~(\ref{eq:form})], and significant values of $|F_l|$ in the optical wavelength range can occur due to the existence of density modulations or charge concentrations at ultrashort length scales. However, following the analytical treatment of microbunching degradation in Ref.~\cite{kim}, we show that the cutoff wavelength in magnetic energy spectrometers is entirely determined by the terms in Eq.~(\ref{eq:longdisp}) with a corresponding strong suppression of coherent emission at optical wavelengths for common magnetic energy spectrometers used at present FELs.
 
		The amount of density modulations in a normalized electron beam distribution $\rho(\vec{X},s)$ with the phase space vector $\vec{X}=(x,x',z,\delta)$ and $\int{d\vec{X}\,\rho(\vec{X},s)}=1$ can be quantified by a complex bunching factor $b(k,s)$ as~\cite{kim}
		\begin{equation}
			b(k,s) = \int{d\vec{X}\,e^{-ikz}\rho(\vec{X},s)}  \,,
		\label{eq:bun}
		\end{equation}
		where $k$ is the wavenumber of the modulation. According to Refs.~\cite{kim,hei}, the evolution of the bunching factor $b[k(s),s]$ along dispersive beamlines can be expressed by
			\begin{equation}
			b[k(s),s] = b_0[k(s),s] + \int_{s_0}^{s}{ds'\,K(s',s)b[k(s'),s']}  \,,
		\label{eq:bu2}
		\end{equation}
		where $b_0[k(s),s]$ is the bunching factor in the absence of collective beam interactions due to CSR. The second term on the right-hand side of the integral equation with the kernel $K(s',s)$~\cite{kim} (a complicated expression that is not relevant here) describes the induced bunching due to CSR interactions. As discussed in Refs.~\cite{kim,csr} and verified by numerical particle tracking simulations below, the bunching induced in a dipole magnet from the energy modulation generated in the same dipole magnet can be neglected with the kernel $K\approx0$, and the bunching factor in Eq.~(\ref{eq:bu2}) becomes $b[k(s),s] \approx b_0[k(s),s]$. This is also the case in a magnetic energy spectrometer consisting of a single dipole magnet, and the resulting evolution of the total bunching factor for a given initial bunching $b_0[k(s_0),s_0]$ can be expressed by~\cite{kim}
		\begin{align}
			b[k(s),s]\approx&\, b_0[k(s_0),s_0]\,\mathrm{exp}\left[ -\frac{k^2(s)\sigma_{\delta0}^2}{2} R_{56}^2\right]\nonumber \\
			          &\times \mathrm{exp}\left[ -\frac{k^2(s)\varepsilon_0\beta_0}{2} \left( R_{51} -\frac{\alpha_0}{\beta_0} R_{52}\right)^2\right]\nonumber \\
				 &\times \mathrm{exp}\left[-\frac{k^2(s)\varepsilon_0}{2\beta_0}R_{52}^2 \right]\,,
		\label{eq:bu3}
		\end{align}
		where the motion in Eq.~(\ref{eq:longdisp}) is taken into account, and an initial beam distribution $\rho(\vec{X}(s_0),s_0)$ that is uniform in $z$ and Gaussian in $x$, $x'$, and $\delta$ is assumed. The initial uncorrelated energy spread and geometrical horizontal emittance are denoted by $\sigma_{\delta0}$ and $\varepsilon_0$, respectively, and $\alpha_0$ and $\beta_0$ are the initial horizontal lattice functions (Twiss parameters). The compression of the wavenumber by $k(s)=k(s_0)[1+hR_{56}(s,s_0)]^{-1}$ with the initial energy chirp $h$ can be neglected, i.e., $k(s)\approx k(s_0)$, since the $R_{56}$ generated by a single dipole magnet is rather small.

		In addition to the evolution of an initial bunching, energy modulations generated upstream of a magnetic energy spectrometer can initiate bunching and, according to Ref.~\cite{kim} and by using Eq.~(\ref{eq:bu3}), the induced bunching $b_E(k,s)$ due to an initial energy modulation is given by

		\begin{equation}
		b_E(k,s) \approx -i k R_{56} \Delta E(k,s_0) \frac{b(k,s)}{b_0(k,s_0)}\,,		
		\label{eq:erg}
		\end{equation}
		where $\Delta E(k,s_0)$ is the Fourier amplitude of the initial energy modulation $\Delta E(z,s_0)$. Fortunately, the bunching $b_E$ can be neglected due to the small $R_{56}$ (see above) and the additional suppression discussed in the following.

		Equation~(\ref{eq:bu3}) implies a suppression of initial bunching due to the coupling with the transverse phase space given in Eq.~(\ref{eq:longdisp}), and a suppression factor $\mathcal{S}$ can be defined as
		\begin{equation}
		\mathcal{S}(k)=\frac{|b(k,s)|^2}{|b_0(k,s_0)|^2}=e^{-k^2\Lambda^2}\,,		
		\label{eq:bu4}
		\end{equation}
		where 
		\begin{equation}
		\Lambda=\sqrt{\varepsilon_0\beta_0\left( R_{51} -\frac{\alpha_0}{\beta_0} R_{52} \right)^2  + \sigma^2_{\delta0}R_{56}^2 +\frac{\varepsilon_0}{\beta_0}R_{52}^2}\,.	
		\label{eq:bu5}
		\end{equation}
		By comparing Eqs.~(\ref{eq:form}) and~(\ref{eq:bun}), and taking into account $\rho[(x,x',z,\delta)]=\rho(z)\rho[(x,x',\delta)]$, the suppression factor can be expressed as $\mathcal{S}(k)=|F_l(k,s)|^2/|F_l(k,s_0)|^2$ (cf. the analytical treatment in Refs.~\cite{bane,rat}), which describes the general suppression of coherent emission in a common magnetic energy spectrometer. Assuming a maximum initial density modulation or an ultrashort electron bunch, both with $|F_l(k,s_0)|\equiv1$, the cutoff wavelength (defined via $|F_l(k_c)| = N^{-1/2}$) is given by [cf. Eq.~(\ref{eq:bu4})]
		\begin{equation}
		\lambda_c = \frac{2\pi\Lambda}{\sqrt{\mathrm{ln}\,N}}\,.
		\label{eq:cut}
		\end{equation}
		We note that the suppression for ultrashort electron bunches is simply given by the lengthening due to the transverse phase space parameters and longitudinal motion given in Eq.~(\ref{eq:longdisp}), which act like a low-pass filter.

			\begin{table}[t]
		\centering
		\caption{Parameters given in the magnetic energy spectrometer at FLASH and used for the particle tracking simulations.}
		\begin{ruledtabular}
		\begin{tabular} {lcccc}
		Parameter           & Symbol       & Value     & Unit  \\ \hline
		Beam energy    &$E$   & 1000    & MeV \\
		Lorentz factor &$\gamma$   & 1957    &  \\
		Electron bunch charge &$Q$& 150 & pC &\\
		Horizontal emittance (normalized)& $\gamma \varepsilon_0$ & 1.0 & $\mu$m \\
		Relative slice energy spread & $\sigma_{\delta0}$ & $10^{-4}$  &  \\
		Horizontal beta function  & $\beta_0$ & 13.55 &  m \\
		Horizontal alpha function & $\alpha_0$ & 5.33 &   \\
		Spatial-to-longitudinal coupling  &$R_{51}$  &   -0.174 & \\
		Angular-to-longitudinal coupling   &$R_{52}$  & -0.089     &  \\
		Momentum compaction factor  &$R_{56}$  &  0.006   & m \\
		\end{tabular}
		\label{tab:spec}
		\end{ruledtabular}
		\end{table}	

		\begin{figure}[b] 
		\centering
		\includegraphics[width=0.9\linewidth]{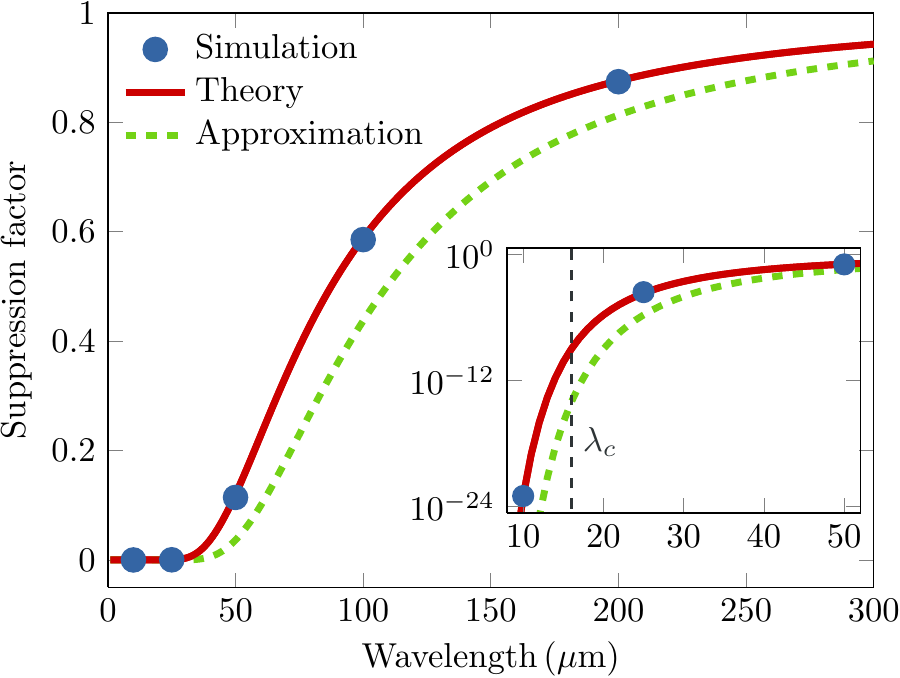}
		\caption{Analytical calculations and numerical simulations (blue dots) of the suppression $\mathcal{S}$ for initial density modulations. The theory curve (solid red line) is calculated for the full term in Eq.~(\ref{eq:bu5}), and the approximation (dashed green line) is calculated for $\Lambda \approx \sqrt{\varepsilon_0\beta_0}  R_{51}$. The inset shows the wavelength range below $52\,\mathrm{\mu m}$ on a logarithmic scale including the cutoff wavelength $\lambda_c$ calculated for $N\approx10^9$ electrons.}
		\label{fig:Supp_4}
		\end{figure}

		The analytical treatment has been verified by numerical simulations using the tracking code {\it elegant}~\cite{Elegant} with Gaussian and uniform beam distributions ($10^6$ particles) including CSR effects, and by using the parameters of the magnetic energy spectrometer at FLASH, summarized in Table~\ref{tab:spec}. Figure~\ref{fig:Supp_4} shows the suppression factor for both numerical simulations with initial density modulations ($10\%$ peak amplitude) and analytical calculations using Eqs.~(\ref{eq:bu4}) and~(\ref{eq:bu5}) for the parameters of FLASH. The analytical calculations are in perfect agreement with the numerical simulations. The shown approximation is calculated by using $\Lambda \approx \sqrt{\varepsilon_0\beta_0}  R_{51}$, which is a good practical estimate ($R_{51} = \sin\Theta$ for a single dipole magnet with bending angle $\Theta$). According to the full term in Eq.~(\ref{eq:bu5}), the cutoff wavelength in the magnetic energy spectrometer at FLASH amounts to $\lambda_c \approx 16\,\mathrm{\mu m}$, which manifests a strong suppression of coherent optical emission.

		\subsection{Suppression of COTR generated by a local ultrashort charge concentration}
		Coherent emission does not only lead to intense radiation, which is described by means of the form factor $|F_l|$ in the intensity distribution given in Eq.~(\ref{eq:spec}), but also to an incorrect representation of the transverse charge distribution in beam profile imaging~\cite{loos}. The imaging of transverse beam distributions with optical systems, e.g., by using an imaging screen, a lens, and a camera, is generally described by means of the intensity distribution of a point source in the image plane (e.g., Ref.~\cite{dao}), which is the so-called point spread function. According to Ref.~\cite{loos}, the image formation with optical transition radiation of a normalized three-dimensional charge distribution $\rho(\vec r,z)$ with $N$ electrons can be expressed by
		\begin{align}
			\left|\vec{\mathcal{E}}(\vec r\,,k)\right|^2=&\,N \int{d\vec{r}\,'dz\,\rho(\vec{r}\,',z)\left|\vec{\mathcal{E}}_1(\vec r- \vec{r}\,',k)\right|^2  }\nonumber \\
			&+ N^2 \left|\int{d\vec{r}\,'dz\,e^{-ikz} \,\rho(\vec{r}\,',z) \vec{\mathcal{E}}_1(\vec r- \vec{r}\,',k)}\right|^2\,,
		\label{eq:field}
		\end{align}
		where $|\vec{\mathcal{E}}(\vec r\,,k)|^2$ describes the measured intensity distribution proportional to the absolute square of the total electric field $\vec{\mathcal{E}}$ evolved from the charge distribution, and $\vec{\mathcal{E}}_1$ corresponds to the imaged electric field of a single electron, which can be expressed by means of the Fresnel-Kirchhoff diffraction integral (e.g., Ref.~\cite{dao}). The second integral in Eq.~(\ref{eq:field}) describes the coherent radiation part ($\sim N^2$), and by taking into account $\rho(\vec r,z)\equiv\rho(\vec r)\rho(z)$ with $\int{d\vec{r}\,\rho(\vec{r}\,)}=\int{dz\,\rho(z)}=1$, the expression for image formation in Eq.~(\ref{eq:field}) can be rewritten as [cf. Eq.~(\ref{eq:spec})]
			\begin{align}
			\left|\vec{\mathcal{E}}(\vec r\,,k)\right|^2=&\,N \int{d\vec{r}\,'\,\rho(\vec{r}\,')\left|\vec{\mathcal{E}}_1(\vec r- \vec{r}\,',k)\right|^2  }\nonumber \\
			&+N^2 \left|F_l(k)\right|^2 \left|\int{d\vec{r}\,'\,\rho(\vec{r}\,') \vec{\mathcal{E}}_1(\vec r- \vec{r}\,',k)}\right|^2\,.	
		\label{eq:field2}
		\end{align}
		The first integral in Eq.~(\ref{eq:field2}) simply describes the incoherent imaging as a convolution of the transverse charge distribution $\rho(\vec{r}\,')$ with the point spread function related term $|\vec{\mathcal{E}}_1|^2$. In the case of a nonvanishing longitudinal form factor $|F_l(k)|\neq0$, the second integral in Eq.~(\ref{eq:field2}) contributes to the image formation and describes no longer a simple convolution with a point spread function, but rather takes into account the actual field distribution. Thus, significant deviations in the measured transverse charge distribution can occur even with a small longitudinal form factor due to the second term $\sim N^2 \left|F_l(k)\right|^2$ in Eq.~(\ref{eq:field2}), where $N\sim 10^9$. An example with initially inconspicuous COTR, impeding the electron beam diagnostics finally, is demonstrated in the following.
		
	\begin{figure}[t]
		\centering
		\subfigure[~K-CCD, $Q\approx0.45\,\mathrm{nC}$.]{\includegraphics[width=0.48\linewidth]{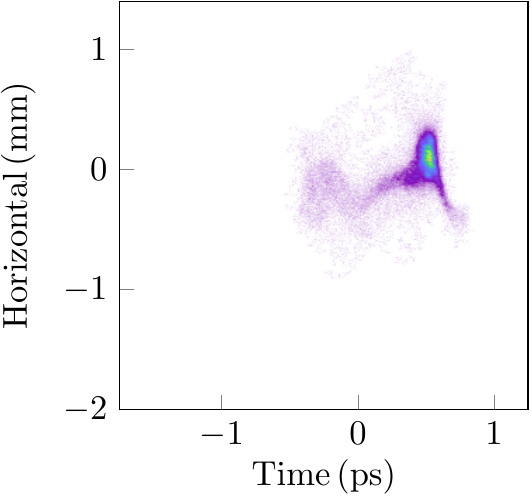} \label{fig:Spike_5_a}}
		\subfigure[~ES-CCD, $Q\approx0.45\,\mathrm{nC}$.]{\includegraphics[width=0.48\linewidth]{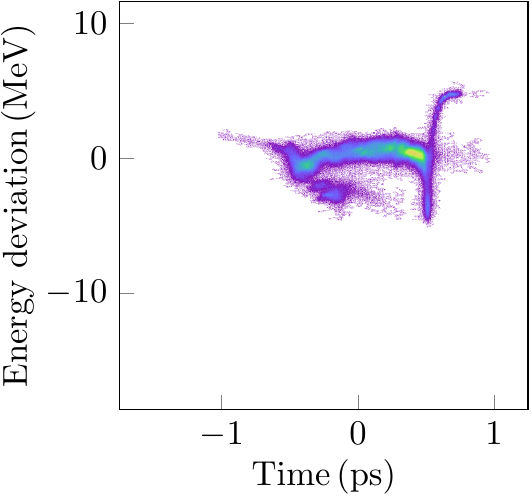} \label{fig:Spike_5_b}}
		\subfigure[~K-CCD, $Q\approx0.55\,\mathrm{nC}$.]{\includegraphics[width=0.48\linewidth]{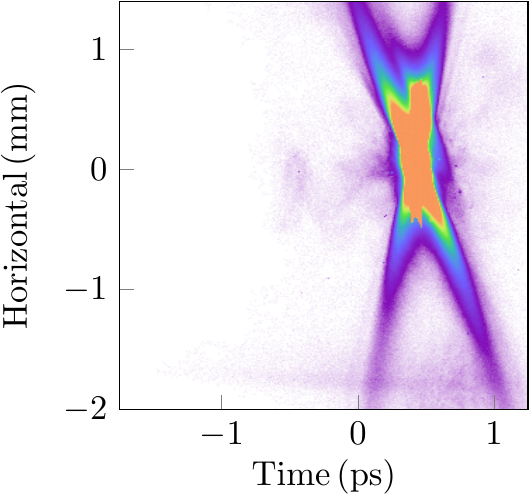} \label{fig:Spike_5_c}}
		\subfigure[~ES-CCD, $Q\approx0.55\,\mathrm{nC}$.]{\includegraphics[width=0.48\linewidth]{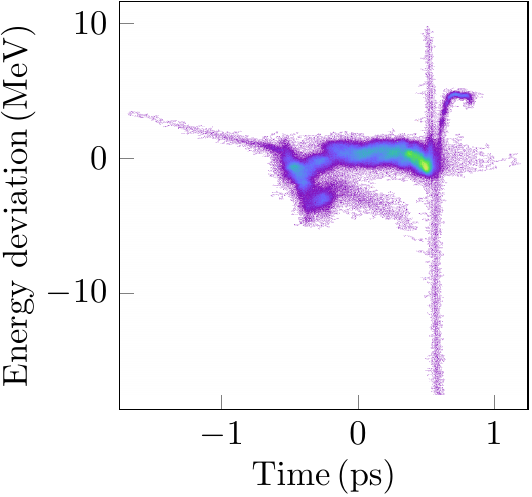} \label{fig:Spike_5_d}}
		\subfigure[~Comparison, $Q\approx0.45\,\mathrm{nC}$.]{\includegraphics[width=0.48\linewidth]{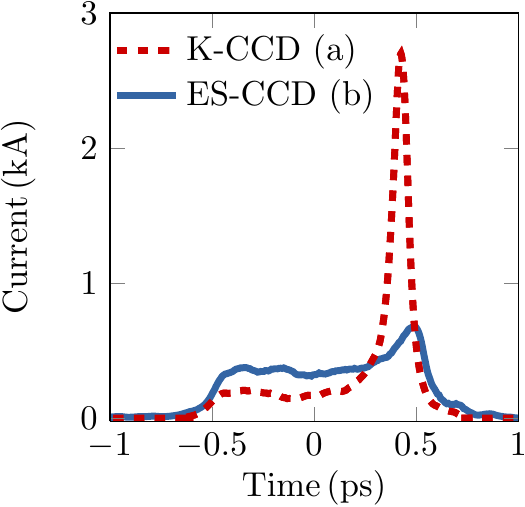} \label{fig:Spike_5_e}}
		\subfigure[~Comparison, $Q\approx0.55\,\mathrm{nC}$.]{\includegraphics[width=0.48\linewidth]{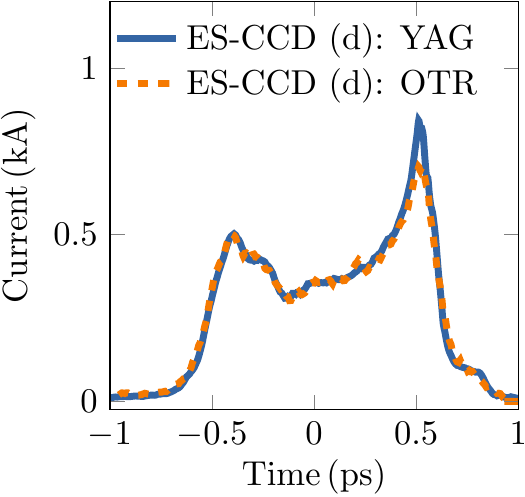} \label{fig:Spike_5_f}}
		\caption{Single-shot measurements of the $t$-$x$ plane in (a) and (c) using a LuAG screen at K-CCD with time $t=-z/c$ (bunch head at $t<0$), and of the longitudinal phase space ($t$,$\Delta E$) in (b) and (d) using a YAG screen at ES-CCD with $\Delta E = \delta E_0$ and $E_0\approx1165\,\mathrm{MeV}$ for bunch charges of $Q\approx0.45\,\mathrm{nC}$ and $0.55\,\mathrm{nC}$, respectively. The comparison of the electron bunch currents between K-CCD and ES-CCD for $Q\approx0.45\,\mathrm{nC}$ is shown in (e), and for $Q\approx0.55\,\mathrm{nC}$ at ES-CCD with different imaging screens it is presented in (f).}
		\label{fig:Spike_5} 
		\end{figure}

	Figures~\ref{fig:Spike_5_a} and~\ref{fig:Spike_5_b} show single-shot images of longitudinal bunch profile measurements using the TDS that were recorded in the non-dispersive main beamline at K-CCD and in the energy spectrometer at ES-CCD, respectively. The images were measured under the same electron beam conditions with a bunch charge of $0.45\,\mathrm{nC}$ and do not display any conspicuous features of COTR. However, as can be seen in Fig.~\ref{fig:Spike_5_e}, the corresponding longitudinal bunch profile taken at K-CCD comprises a much narrower spike with higher peak current. When increasing the bunch charge to $0.55\,\mathrm{nC}$, COTR emission became apparent at K-CCD [Fig.~\ref{fig:Spike_5_c}], whereas the image in the energy spectrometer at ES-CCD [see Fig.~\ref{fig:Spike_5_d}] did not show any coherent radiation effects. The COTR emission in Fig.~\ref{fig:Spike_5_c} (we chose a single-shot image with low saturation of the CCD) is clearly localized in the longitudinal electron bunch profile at a time coordinate of about 0.5\,ps. At the same time coordinate, the longitudinal phase space in Fig.~\ref{fig:Spike_5_d} exhibits a huge but narrow increase in energy spread (the width in the time is limited by the TDS resolution). From this we conclude that the single-shot image in Fig.~\ref{fig:Spike_5_a} already partially contains COTR as a consequence of a small but nonvanishing form factor $|F_l|$ [cf. Eqs.~(\ref{eq:spec}) and~(\ref{eq:field2})] and that the COTR emission in Fig.~\ref{fig:Spike_5_c} seems most probably to be generated by a local ultrashort charge concentration such as a sharp spike inside the electron bunch. We note that the measurements presented in Fig.~\ref{fig:Spike_5_e} should give the same longitudinal electron bunch profiles, and the existing deviations cannot be explained due to a worse resolution as is the case in Sec.~\ref{sec:lp}. In order to demonstrate the local energy spread increase in Figs.~\ref{fig:Spike_5_b} and~\ref{fig:Spike_5_d} with a reasonable signal-to-noise ratio (SNR), the longitudinal phase space measurements are presented with the YAG imaging screen. The measurement performed with the OTR imaging screen, presented in Fig.~\ref{fig:Spike_5_f}, shows the same strong COTR suppression (but worse SNR).

	\section{Techniques for separation of coherent optical radiation}\label{sec:sep}

As demonstrated in Sec.~\ref{sec:ES}, electron beam profile measurements can be accomplished in dispersive beamlines, such as magnetic energy spectrometers, with standard optical imaging systems as the emission of coherent optical radiation is strongly suppressed. However, linear accelerators consist mainly of beamlines which are in general designed to be dispersion-free, and imaging in energy spectrometers precludes measuring pure transverse beam profiles due to the dispersion. In this section, we discuss methods that suppress the impact of coherent radiation by separation from an incoherent radiation part.

		\subsection{Spectral separation}\label{sec:sepspec}
		\begin{figure}[t] 
		\centering
		\includegraphics[width=0.9\linewidth]{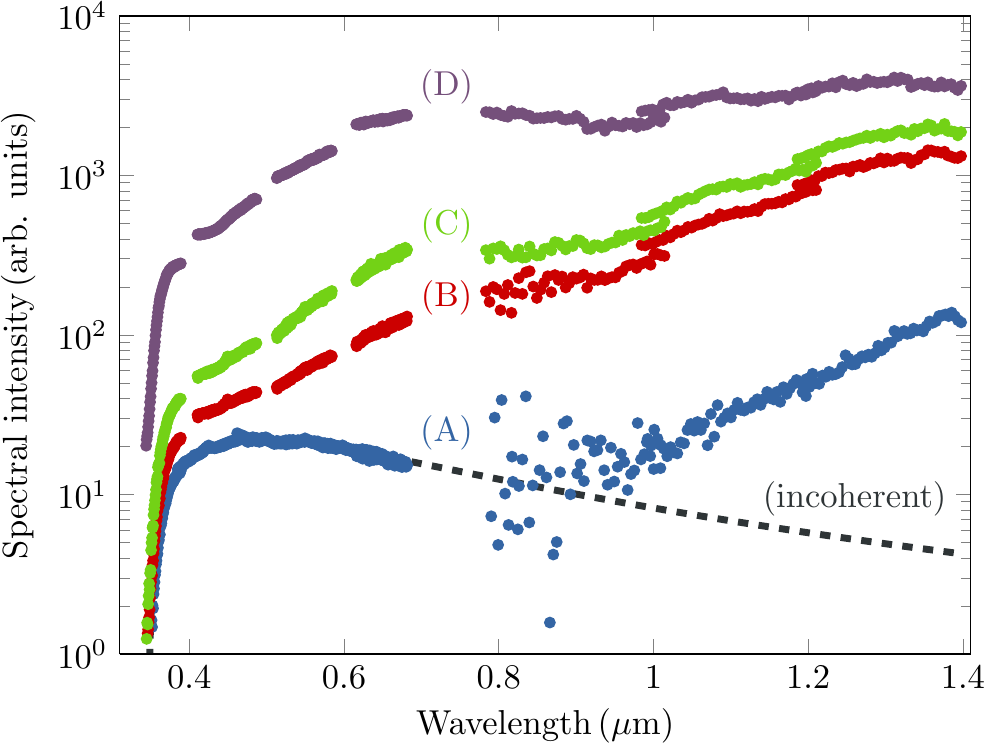}
		\caption{Spectral intensity measurements of transition radiation in the visible and near-infrared wavelength range for four different compression settings: (A) FEL operation and (B) - (D) marginal compression, i.e., on-crest rf operation with decreasing $R_{56}$ in the bunch compressors (see Ref.~\cite{Wesch1} for experimental details). The spectral intensity of the incoherent part of transition radiation is indicated as dashed black line.}
		\label{fig:Spec_6}
		\end{figure}
The spectral intensity of transition (synchrotron) radiation emitted by an electron bunch consists of two terms that describe the incoherent ($\sim N$) and coherent ($\sim N^2|F_l|^2|F_t|^2$) radiation part [cf.~Eq.~(\ref{eq:spec}) or Eq.~(\ref{eq:field2})]. A spectral separation of these terms in electron beam profile imaging can be accomplished by restricting the imaging with wavelengths below the cutoff wavelength $\lambda_{c}$, i.e., where the emission is dominated by incoherent radiation. Spectral separation has been considered in Ref.~\cite{lump2} by using a scintillation screen in combination with a bandpass filter. However, this method requires a good knowledge and control of the expected spectra, and a vanishing form factor ($|F_l||F_t| \ll N^{-1/2}$) in the detectable wavelength range, which is not the general case as the spectra can vary strongly with the operation modes of a linear accelerator. This is demonstrated in Fig.~\ref{fig:Spec_6}, in which spectral measurements of transition radiation in the visible and near-infrared wavelength range are presented for different compression settings at FLASH. The dashed black line represents the incoherent radiation part convoluted with the transmission of the optical setup. In contrast to the measurements presented in Sec.~\ref{sec:subsecCOTR}, the measurements shown in Fig.~\ref{fig:Spec_6} were performed upstream of the collimator section. We note that similar, reproducible measurements for uncompressed electron bunches, showing coherent radiation prominently at the micrometer scale, have been presented in Ref.~\cite{Wesch2}, and COTR for uncompressed bunches has been reported in Ref.~\cite{loos}.

In general, the probability of coherent emission decreases at shorter wavelengths, which is often not sufficiently reduced for optical wavelengths, and imaging with transition radiation in the EUV region might be an option~\cite{suk,gero,gero2}. In addition to the knowledge and control of the spectra, the imaging with EUV radiation also requires dedicated detectors and optics, and a complete set-up in vacuum to prevent strong absorption in air. 

	\begin{figure*}[htb]
		\centering
		\subfigure[~OTR screen, no time delay.]{\includegraphics[width=0.3\linewidth]{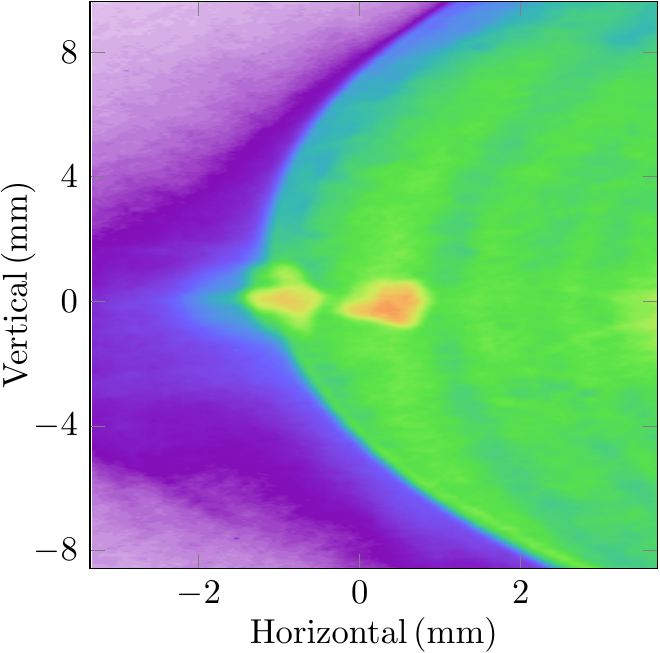} \label{fig:Proof_7_a}}
		\subfigure[~LuAG screen, no time delay.]{\includegraphics[width=0.3\linewidth]{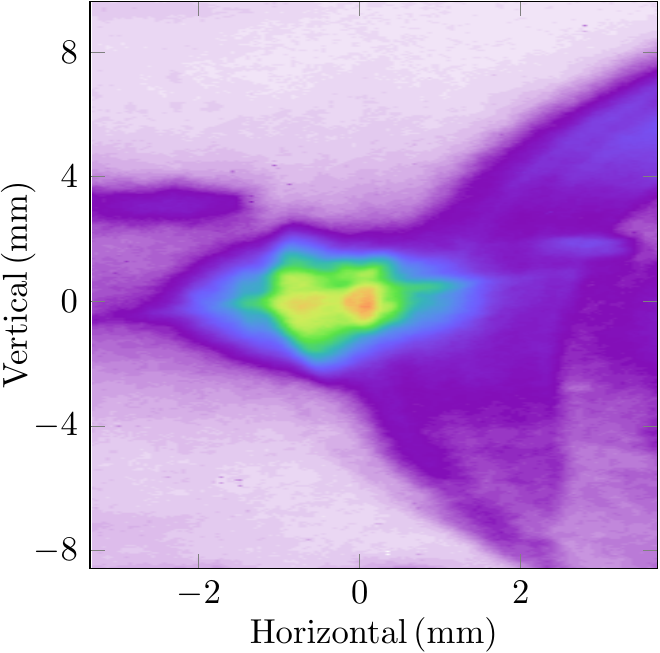} \label{fig:Proof_7_b}}
		\subfigure[~LuAG screen, time delay.]{\includegraphics[width=0.3\linewidth]{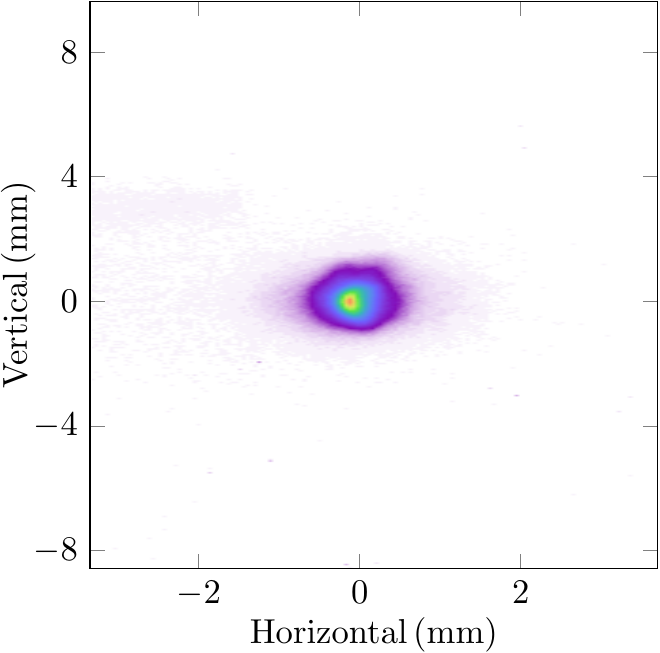} \label{fig:Proof_7_c}}
		\caption{Proof-of-principle for the temporal separation technique in transverse beam profile imaging, demonstrated for compressed electron bunches at K-ICCD with the three screen/readout configurations: (a) OTR screen, (b) LuAG screen, and (c) LuAG screen with delayed readout. The images in (a) and (b) show a composite of optical transition and synchrotron radiation with a contribution of scintillation light in (b). The image in (c) is expected to show delayed but pure scintillation light.}
		\label{fig:Proof_7}
		\end{figure*}
		\subsection{Spatial separation}
		The luminescence of scintillation screens~\cite{lecog}, which is a stochastic process, is inherently linear in the number of interacting electrons (neglecting quenching and saturation effects), hence coherent radiation effects are not expected in pure scintillation light. However, transition radiation is also emitted at the boundary of vacuum and scintillator, and coherent optical radiation can still appear [see, e.g., Fig.~\ref{fig:Indi_3_b}]. Then, the total spectral and angular intensity distribution can be written as (omitting the arguments $(k,\Omega)$ in the intensity distributions $\mathcal{I}$)
		\begin{equation}
			\mathcal{I}_t = N \mathcal{I}_s  + \left[N + N^2 |F_l(k)|^2|F_t(k,\Omega)|^2\right]\mathcal{I}_o\,,
		\label{eq:specFull}
		\end{equation} 
		where $\mathcal{I}_s$ and $\mathcal{I}_o$ are related to scintillation light and transition radiation, respectively. As discussed in Sec.~\ref{sec:sepspec} for OTR imaging screens and with the same requirements and restrictions, spectral separation can also be applied when using scintillation screens ($\mathcal{I}_t \approx N \mathcal{I}_s  + N \mathcal{I}_o$). Another method, particularly suited for scintillation screens, which have nearly isotropic emission, is to make use of the strong angular dependence of optical transition radiation (e.g., Refs.~\cite{hon,dao}) and to perform electron beam profile imaging with radiation that is dominated by scintillation light, i.e., $ \mathcal{I}_o(k,\Omega) \ll \mathcal{I}_s(k,\Omega)/[1+N|F_l(k)|^2|F_t(k,\Omega)|^2] $ in Eq.~(\ref{eq:specFull}). Spatial separation can be achieved with imaging geometries having large angular or spatial offsets, e.g., by using tilted imaging screens~\cite{yan1} or central masks~\cite{sp8}, where $\mathcal{I}_o(\Omega)|F_t(\Omega)|^2$ is suppressed sufficiently. However, just as for spectral separation, this method also requires good knowledge and control of the form factor, and dedicated imaging geometries. In addition, the resolution depends on the observation angle of the scintillation screen (e.g., Ref.~\cite{yan1}), which has to be taken into account in the layout of the imaging system. We note that an experiment on the spatial separation technique is currently being commissioned at FLASH.

		\subsection{Temporal separation}\label{sec:sept}
		The fundamentally different light generation processes of scintillators and optical transition radiators result in clearly distinct temporal responses. The emission of transition radiation from relativistic electrons is instantaneous ($\sim\mathrm{fs}$) and prompt~\cite{lump3,mary} compared to the decay times ($\sim\mathrm{ns}$) of common scintillators (e.g., Ref.~\cite{lecog}). Accordingly, the temporal profiles of the OTR pulses resemble the longitudinal electron beam profiles, whereas the temporal scintillation light pulses are fully dominated by the decay of the excited states in the scintillator. Temporal separation makes use of the distinct temporal responses and allows to entirely eliminate OTR, i.e., the term $\mathcal{I}_o$ in Eq.~(\ref{eq:specFull}) which is time-dependent with $\mathcal{I}_o\equiv\mathcal{I}_o(k,\Omega,t)$, and, therewith, coherent optical radiation effects in electron beam profile imaging with scintillation screens when reading out a gated camera with a certain time delay after the prompt emission of OTR. Image recording with delayed readout (e.g., Ref.~\cite{mary}) can be accomplished with intensified CCD (ICCD) cameras, where a control voltage in the intensifier between photocathode and micro-channel plate allows fast gating and exposure times of a few nanoseconds (e.g., Refs.~\cite{lump3,mary,ga1}). The experiments on the temporal separation technique at FLASH have been performed by using the ICCD camera ``PCO: Dicam Pro (S20)''~\cite{pco} in combination with the off-axis LuAG scintillation imaging screen in the non-dispersive main beamline at K-ICCD, which has a decay time of $\sim50\,\mathrm{ns}$~\cite{bla}. The cameras used for the presented measurements are able to readout images at the bunch train repetition rate of FLASH of 10\,Hz, hence one bunch per bunch train can be measured with single-shot capability. Further technical details on the equipment used for the measurements presented in the following can be found in Sec.~\ref{sec:screens} and in Refs.~\cite{yan1,yan2}.

The series of single-shot images in Fig.~\ref{fig:Proof_7} present first proof-of-principle measurements on the temporal separation technique. The image shown in Fig.~\ref{fig:Proof_7_a} was recorded at K-ICCD with an OTR screen, whereas for Figs.~\ref{fig:Proof_7_b} and~\ref{fig:Proof_7_c} a LuAG scintillation screen was used. The image shown in Fig.~\ref{fig:Proof_7_c} has been recorded with a time delay of $100\,\mathrm{ns}$, which is rather long compared to the emission time of OTR but takes into account the large camera trigger-jitter that existed during the measurements. The image recorded with the OTR screen and time delay simply showed background noise and is not presented here. The intensity distributions in Fig.~\ref{fig:Proof_7} have been generated by moderately compressed electron bunches with a charge of 0.5\,nC and a beam energy of 700\,MeV. Figures~\ref{fig:Proof_7_a} and~\ref{fig:Proof_7_b} show a composite of COTR and COSR with a contribution of scintillation light in Fig.~\ref{fig:Proof_7_b}. The round-shaped light pattern on the right-hand side of Figs.~\ref{fig:Proof_7_a} and~\ref{fig:Proof_7_b} is most probably due to synchrotron radiation generated upstream of the off-axis screens (a polarizer was not available during the measurements), where the appearance in Fig.~\ref{fig:Proof_7_b} is reduced by the transparency of the LuAG screen. The image in Fig.~\ref{fig:Proof_7_c}, recorded with a time delay of $100\,\mathrm{ns}$, can be attributed purely to scintillation light allowing for a quantitative analysis of the transverse beam profiles.	
	
In contrast to spectral and spatial separation, the temporal separation technique provides a definite method to suppress coherent optical transition radiation without further relying on the wavelength-dependent longitudinal form factor. In addition, this technique inherently includes the suppression of secondary incoherent radiation sources such as synchrotron radiation generated from magnets directly upstream of the imaging screen or backward OTR emitted from the second imaging screen boundary, whereas spectral components in the UV region or at shorter wavelengths may excite the scintillator, affecting the temporal separation. As is shown in Ref.~\cite{bane}, however, potential synchrotron radiation sources can be identified and thus separated by adjusting the upstream magnets. Furthermore, the coherent emission of OTR at the second scintillator screen boundary is mitigated due to multiple scattering in the scintillator material as is described and demonstrated in Refs.~\cite{loos,scatter}. We note that the current implementation of the temporal separation technique presented throughout this paper utilizes fast ICCD cameras, which are currently an order of magnitude more expensive than conventional CCD cameras.

	\section{Experimental results with temporal separation}\label{sec:res}
	The proof-of-principle measurements on the temporal separation technique presented in Fig.~\ref{fig:Proof_7} were carried out at K-ICCD. However, a reference measurement to quantitatively prove this technique in terms of transverse beam profiles, as would be provided by a wire-scanner, which is insensitive to coherent effects, is not available at this position. In this section, we verify the method of temporal separation by investigations on the charge-dependent image intensities and comparisons with longitudinal bunch profiles recorded in the energy spectrometer at ES-CCD.

	\subsection{Charge dependence of integrated intensity}
		\begin{figure}[t] 
		\centering
		\includegraphics[width=0.99\linewidth]{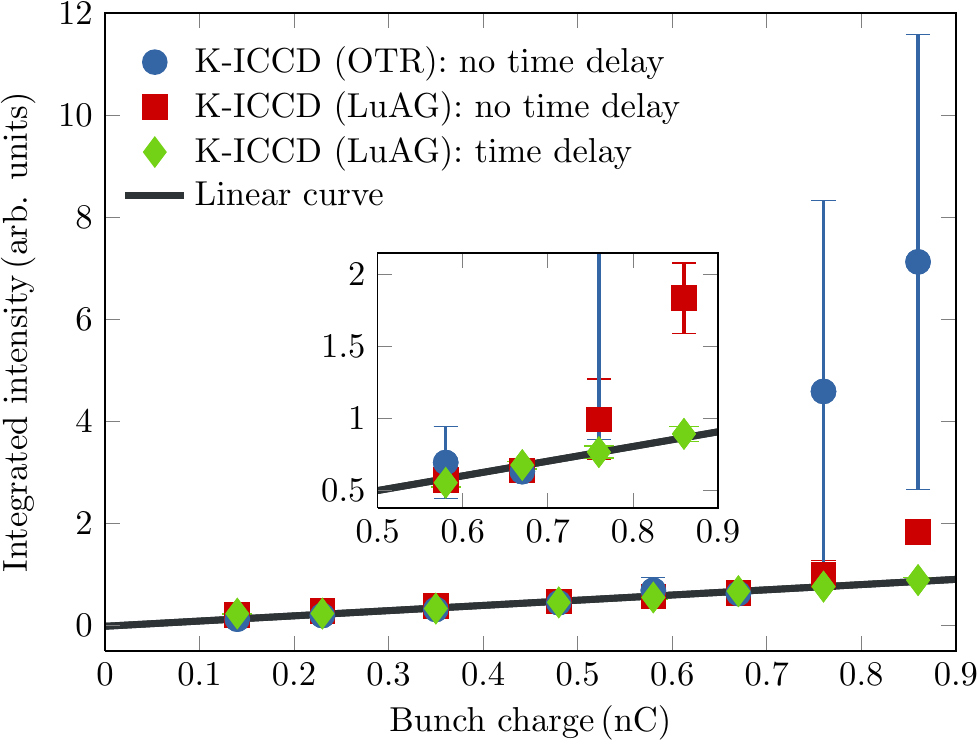}
		\caption{Measurements on the bunch charge dependence of the integrated intensity at K-ICCD generated by compressed electron bunches using different screen/readout configurations, where the inset shows the range from 0.55\,nC to 0.9\,nC. The linear curve shows the dependence of incoherent radiation.}
		\label{fig:Charge_8}
		\end{figure}
		
		Incoherent radiation is linear in the number of electrons contributing to the emission process (cf. Sec.~\ref{sec:supp}), i.e., linear in the electron bunch charge ($\sim Q$), and deviations caused by the nonlinear charge dependence of coherent radiation ($\sim |F_l|^2Q^2$) are ideally suited to verify the temporal separation technique. The integrated image intensities presented in Fig.~\ref{fig:Charge_8} were measured for bunch charges between 0.13\,nC and 0.87\,nC at K-ICCD for different imaging screen and readout configurations. Each data point represents the average intensity of 20 background-corrected single-shot images and the error bars indicate the statistical r.m.s.~image intensity fluctuations. Up to an electron bunch charge of $Q\sim 0.5\,\mathrm{nC}$, the integrated intensity is linear (solid black line) in $Q$ for all presented configurations. For higher bunch charges, deviations from the linear dependence appear in the configurations without delayed readout, i.e., the form factor $|F_l|$ becomes significant in the visible wavelength range, which are caused by contributions from coherent optical radiation. The inset in Fig.~\ref{fig:Charge_8} shows the bunch charge range from 0.55\,nC to 0.9\,nC more detailed. We note that the integrated intensity of the OTR (blue dots) has actually been higher than presented for $Q>0.7\,\mathrm{nC}$, because of camera saturation due to the strong optical emission and the corresponding underestimated integrated intensity. The large error bars, representing the r.m.s.~jitter, indicate strong fluctuations due to the COTR. In the case of the LuAG imaging screen recorded with a time delay (green diamonds), the dependence of the integrated intensity is entirely linear in the bunch charge, which verifies the power of the temporal separation technique.

		\subsection{Longitudinal electron bunch compression}\label{sec:ref}
		\begin{figure}[t] 
		\centering
		\includegraphics[width=1\linewidth]{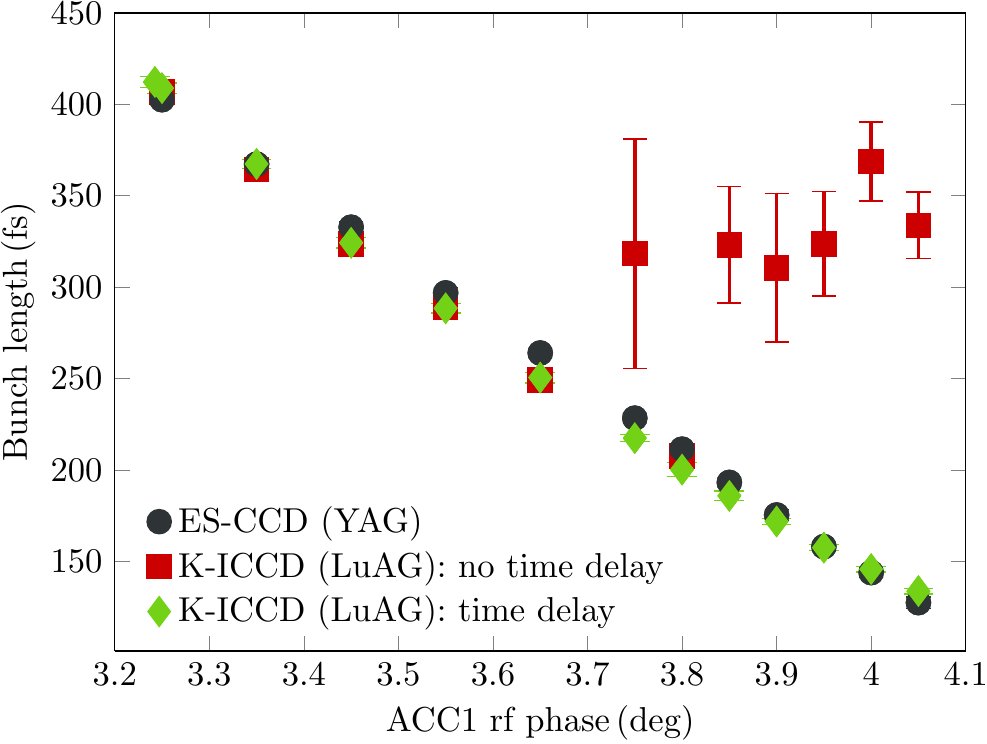}
		\caption{Electron bunch length measurements for varying ACC1 rf phase at ES-CCD and K-ICCD using different readout configurations. According to Sec.~\ref{sec:ES}, the measurements in the magnetic energy spectrometer (``ES-CCD (YAG)'') are intended to provide an absolute reference measurement.}
		\label{fig:Compression_9}
		\end{figure}
	
		As the emission of COTR is strongly suppressed in the magnetic energy spectrometer at FLASH (see Sec.~\ref{sec:ES}), electron bunch profiles measured at the screen station ES-CCD can serve as a reference for comparison with the temporal separation technique applied in the non-dispersive beamline at K-ICCD. While the transverse bunch profiles can differ at both locations due to different Twiss parameters and dispersion at ES-CCD, longitudinal bunch compression does not take place in between, and longitudinal bunch profile measurements using the TDS can be used for a direct comparison. The measurements presented in Fig.~\ref{fig:Compression_9} show the mean r.m.s.~electron bunch length of 20 single-shot images, including the statistical r.m.s.~jitter indicated via error bars, for various ACC1 rf phases measured at ES-CCD and K-ICCD by using the TDS. The electron bunches were set up with an energy of 700\,MeV and a bunch charge of 0.5\,nC. The rf phase of ACC1 affects the energy chirp of the electron bunches upstream of the first bunch compressor and, accordingly, the final electron bunch lengths. The r.m.s.~electron bunch lengths measured in the magnetic energy spectrometer at ES-CCD (black dots) decrease almost linearly and do not possess large fluctuations.
	
		In contrast to the magnetic energy spectrometer at ES-CCD, coherent optical emission is not suppressed in the non-dispersive beamline at K-ICCD, leading to a sudden increase of the r.m.s.~electron bunch lengths in combination with large fluctuations, represented by the large error bars (statistical r.m.s.~jitter), for ACC1 rf phases $\gtrsim3.75\,\mathrm{deg}$ measured with a LuAG screen without a certain time delay (red squares), i.e., without applied temporal separation. The electron bunch length measurements using an OTR screen are omitted in Fig.~\ref{fig:Compression_9} due to even larger deviations and fluctuations compared to the reference at ES-CCD for ACC1 rf phases $\gtrsim3.75\,\mathrm{deg}$. Instead, the OTR images (single-shots) for ACC1 rf phases of 3.25\,deg and 3.75\,deg are presented in Figs.~\ref{fig:Comp_10_a} and~\ref{fig:Comp_10_b}, respectively, with obvious coherent optical radiation effects in Fig.~\ref{fig:Comp_10_b}. Due to the fact that the electron beam images shown in Fig.~\ref{fig:Comp_10} are sheared vertically by means of the TDS, the vertical coordinate implies time information (see Eq.~\ref{eq:motion}) and the faint bunching visible in Fig.~\ref{fig:Comp_10_a} may be assigned to microbunching. 
				
		Figure~\ref{fig:Comp_10_c} shows a single-shot image taken at K-ICCD using a LuAG screen without time delay for an ACC1 rf phase of 3.75\,deg. The image clearly shows contributions of coherent optical radiation similar to the image in Fig.~\ref{fig:Comp_10_b}. By imaging the LuAG screen with a time delay of 100\,ns, the obtained distribution shown in Fig.~\ref{fig:Comp_10_d} is acceptable without obvious contributions from coherent optical radiation. In addition, the corresponding electron bunch length measurements with applied temporal separation (green diamonds) in Fig.~\ref{fig:Compression_9} are in perfect agreement with the reference measurements in the energy spectrometer at ES-CCD (black dots). The electron bunch durations for FEL operation at FLASH are typically shorter than 150\,fs (e.g., Ref.~\cite{pulse}), and typical electron beam parameters are given in Table~\ref{tab:spec}.

		\subsection{Longitudinal electron beam profiles}\label{sec:lp}
		\begin{figure*}[t]
		\centering
		\subfigure[~3.25\,deg, OTR.]{\includegraphics[width=0.239\linewidth]{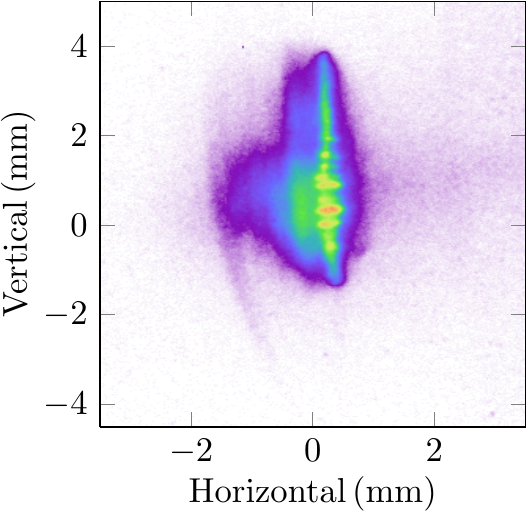} \label{fig:Comp_10_a}}
		\subfigure[~3.75\,deg, OTR.]{\includegraphics[width=0.239\linewidth]{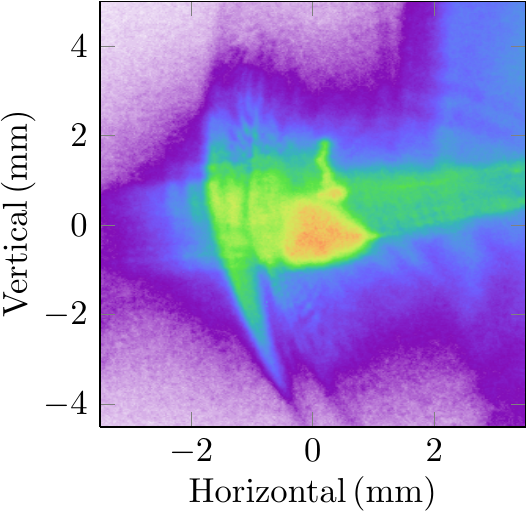} \label{fig:Comp_10_b}}
		\subfigure[~3.75\,deg, LuAG.]{\includegraphics[width=0.239\linewidth]{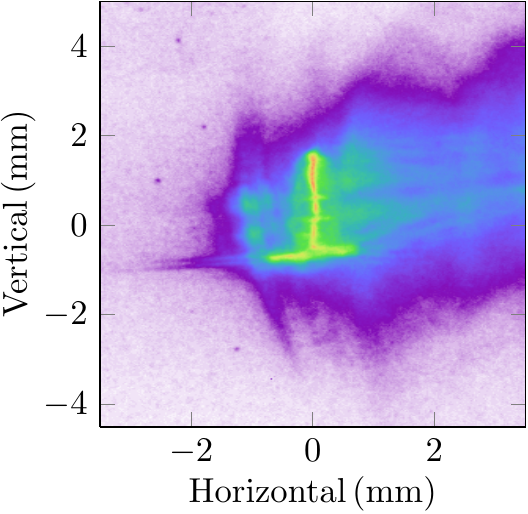} \label{fig:Comp_10_c}}
		\subfigure[~3.75\,deg, LuAG, time delay.]{\includegraphics[width=0.239\linewidth]{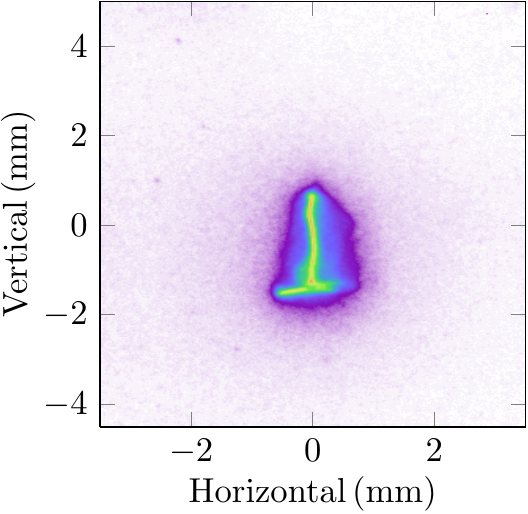} \label{fig:Comp_10_d}}
		\caption{Single-shot electron beam profile images at K-ICCD for two ACC1 rf phases used in the measurements shown in Fig.~\ref{fig:Compression_9} with different screen/readout configurations: (a) OTR screen for 3.25\,deg, (b) OTR screen for 3.75\,deg, and (c) LuAG screen for 3.75\,deg without and (d) with delayed readout. The presented single-shot beam profile images are background-corrected.}
		\label{fig:Comp_10}
		\end{figure*}

		\begin{figure}[t]
		\centering
		\subfigure[~ACC1 rf phase of 3.75\,deg.]{\includegraphics[width=0.48\linewidth]{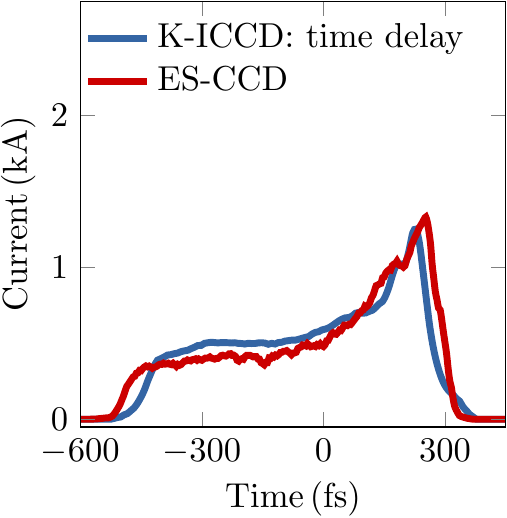} \label{fig:Prof_11_a}}
		\subfigure[~ACC1 rf phase of 4.05\,deg.]{\includegraphics[width=0.48\linewidth]{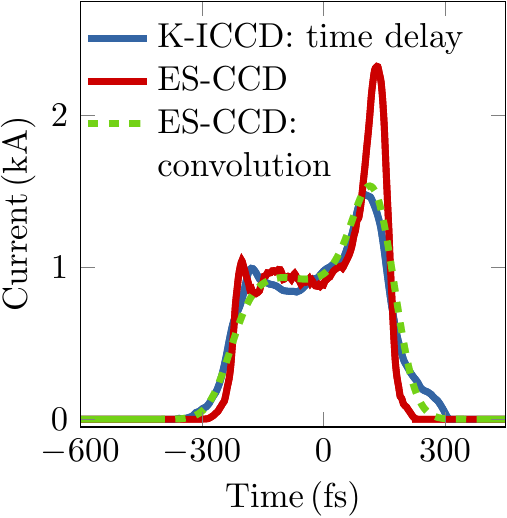} \label{fig:Prof_11_b}}
		\caption{Single-shot longitudinal electron bunch profiles measured in the non-dispersive beamline at K-ICCD by using the LuAG screen recorded with a certain time delay (blue line) and in the magnetic energy spectrometer at ES-CCD by using the YAG screen (red line) for ACC1 rf phases of (a) 3.75\,deg and (b) 4.05\,deg, respectively. A convolution has been applied in (b) for the measurement at ES-CCD (green dashed line) to compare the longitudinal profiles with similar resolution. }
		\label{fig:Prof_11}
		\end{figure}	
		The temporal separation technique, which has demonstrated accurate r.m.s.~electron bunch length measurements in the presence of coherent optical radiation effects, gives confidence that single-shot measurements of longitudinal bunch profiles and, accordingly, electron bunch currents using temporal separation result in reliable results. The single-shot measurements presented in Fig.~\ref{fig:Prof_11} (cf. measurements shown in Figs.~\ref{fig:Compression_9} and~\ref{fig:Comp_10} for the same ACC1 rf phase settings) have been recorded for an ACC1 rf phases of 3.75\,deg in Fig.~\ref{fig:Prof_11_a} and for 4.05\,deg in Fig.~\ref{fig:Prof_11_b}, i.e., in the presence of coherent optical radiation effects. The longitudinal electron bunch profiles taken in the non-dispersive beamline at K-ICCD (blue line) with temporal separation show good agreement with the reference measurements at ES-CCD (red line), and the observed deviations are most probably due to slightly nonlinear amplification in the intensifier (photocathode and micro-channel plate) of the ICCD camera. The reduced peak current with broadening in time in the case of ``K-ICCD: time delay``, which is apparent on the right-hand side ($\mathrm{Time}>0$) of Fig.~\ref{fig:Prof_11_b}, can be explained by the different time resolutions of $\mathcal{R}_{t,\mathrm{e}}=13\,\mathrm{fs}$ and $43\,\mathrm{fs}$ achieved with the TDS during the measurements for ES-CCD and K-ICCD, respectively. In order to compare the longitudinal bunch profiles with comparable resolution, a convolution has been applied for the measurement at ES-CCD in Fig.~\ref{fig:Prof_11_b} by taking into account the actual time resolution. The longitudinal bunch profile after carrying out the convolution (green dashed line) is in good agreement with the bunch profile taken at K-ICCD with applied temporal separation (blue line).

	\section{Summary and conclusions}\label{sec:Summary}
	Electron beam profile imaging is crucial for many applications in electron beam diagnostics at FELs, and particularly required to perform single-shot diagnostics. However, the frequent appearance of coherent optical radiation effects, e.g., COTR, in high-brightness electron beams impedes incoherent beam profile imaging with standard techniques. The theoretical considerations, numerical simulations, and experimental data presented in this paper show that coherent optical emission can be strongly suppressed by performing beam profile imaging in a magnetic energy spectrometer due to sufficient spatial-to-longitudinal coupling. However, energy spectrometers preclude measuring pure transverse beam profiles due to dispersion in the bending plane. For incoherent beam profile imaging in non-dispersive beamlines, we discussed methods to separate the incoherent radiation from scintillation screens and to simultaneously exclude coherent optical radiation from detection. In contrast to spectral and spatial separation, the temporal separation technique, utilizing an ICCD camera, provides a definite method to suppress coherent optical transition radiation without knowledge of the longitudinal form factor. In terms of readout times and rates, ICCD cameras have the same applicability as standard CCD cameras. By applying the temporal separation technique in the presence of coherent optical radiation, we demonstrated reliable measurements of longitudinal electron beam profiles, and measurements of r.m.s.~electron bunch lengths in excellent agreement with reference measurements in a magnetic energy spectrometer. Limitations may appear due to scintillator excitation by secondary coherent radiation sources. However, the presented experimental results prove the temporal separation technique as a promising method for future applications in beam profile diagnostics for high-brightness electron beams.

	\begin{acknowledgments}
	We would like to thank the whole FLASH-team, and the engineers and technicians of the DESY groups FLA, MCS, and MVS for their great support. We also thank B. Faatz, K. Honkavaara, and S. Schreiber for providing beam time, and Y. Ding and H. Loos for fruitful discussions. In particular, we are deeply grateful to E.A. Schneidmiller for careful reading of the manuscript and to Z. Huang for providing many helpful explanations. 

	\end{acknowledgments}

\end{document}